\documentclass[12pt,preprint]{aastex63}

\usepackage{graphicx}
\usepackage{natbib}
\usepackage{color}
\usepackage{amssymb}
\usepackage{amsthm}
\usepackage{amsmath}
\usepackage{bm}
\usepackage{float}
\usepackage{placeins}
\usepackage{color,soul}
\usepackage{hyperref}
\usepackage{tabularx}
\usepackage{subfigure}

\graphicspath{{./}{figures/}}

\begin{document}

\title{Decreasing False Alarm Rates in ML-based Solar Flare Prediction using SDO/HMI Data.}

\author{Varad Deshmukh}
\affil{Dept. of Computer Science, University of Colorado Boulder, Boulder, CO, 80309 USA}
\email{varad.deshmukh@colorado.edu}

\author{Natasha Flyer}
\affil{Flyer Research LLC, Boulder, CO 80303 USA}
\email{natasha.flyer@gmail.com}

\author{Kiera van der Sande}
\affil{Dept. of Applied Mathematics, University of Colorado Boulder, Boulder, CO, 80309 USA}
\email{kiera.vandersande@colorado.edu}

\author{Thomas Berger}
\affil{Space Weather Technology, Research, and Education Center, University of Colorado at Boulder \\
3775 Discovery Drive, Boulder, CO 80303}
\email{thomas.berger@colorado.edu}

\begin{abstract}
A hybrid two-stage machine learning architecture that addresses the problem of excessive false positives (false alarms) in solar flare prediction systems is investigated. The first stage is a convolutional neural network (CNN) model based on the VGG-16 architecture that extracts features from a temporal stack of consecutive Solar Dynamics Observatory (SDO) Helioseismic and Magnetic Imager (HMI) magnetogram images to produce a flaring probability. The probability of flaring is added to a feature vector derived from the magnetograms to train an extremely randomized trees (ERT) model in the second stage to produce a binary deterministic prediction (flare/no flare) in a 12-hour forecast window. To tune the hyperparameters of the architecture a new evaluation metric is introduced, the ``scaled True Skill Statistic". It specifically addresses the large discrepancy between the true positive rate and the false positive rate in the highly unbalanced solar flare event training datasets. Through hyperparameter tuning to maximize this new metric, our two-stage architecture drastically reduces false positives by $\approx$ $48\%$ without significantly affecting the true positives (reduction by $\approx$ $12\%$), when compared with predictions from the first stage CNN alone. This, in turn, improves various traditional binary classification metrics sensitive to false positives such as the  precision, F1 and the Heidke Skill Score. The end result is a more robust 12-hour flare prediction system that could be combined with current operational flare forecasting methods. Additionally, using the ERT-based feature ranking mechanism, we show that the CNN output probability is highly ranked in terms of flare prediction relevance.  
\end{abstract}



\section{Introduction} 
\label{sec:intro}
Solar flares are the electromagnetic radiation outbursts accompanying Solar Magnetic Eruptions (SMEs) in the outer solar atmosphere \citep{Fletcher:2011}. The resulting X-ray and EUV radiation ionizes Earth's upper atmosphere and can cause radio, radar, and Global Navigation Satellite System (GNSS) signal interference on the sunlit side of the Earth. In addition to flares, SMEs can also result in ``Coronal Mass Ejections'' (CMEs),  large magnetic plasma clouds launched into interplanetary space at speeds on the order of $10^3$~$\mathrm{km~s^{-1}}$ \citep{Webb:2012}. If CMEs impact the Earth's magnetosphere, they can trigger geomagnetic storms which can produce a variety of impacts including large-scale inner magnetospheric currents, global ionospheric disturbances, increased drag on satellites in Low-Earth orbit, and, in more severe cases, geoelectric fields that can destabilize or damage electric power transmission grids \citep{Lucas:2020}. SMEs and the shock waves propagating in front of the associated CMEs can also accelerate charged particles to relativistic energies, resulting in ``radiation storms'' that propagate along the heliospheric magnetic field \citep{Reames:2013} to endanger astronauts in space and potentially damage spacecraft avionics. Historically, solar flares were the first phenomenon discovered to be associated with SMEs \citep[e.g.,][]{Carrington:1859} and the electromagnetic radiation from flares is the first indication we receive of an eruption on the Sun. Indeed, the size of SMEs is still generally described by the intensity of the X-ray irradiance from the associated flare. Thus while the ultimate goal is to predict SMEs, we follow common usage and apply the term ``solar flare prediction'' to describe the goal of our investigation. 

Currently, operational solar flare prediction is accomplished using manual classification of sunspot active region (AR) shape, size, and complexity in visible-light images of the solar photosphere. The classifications are then associated with historical 24-, 48-, and 72-hour probabilities of producing X-ray flares of a given magnitude via look-up tables \citep{McIntosh:1990wu} and modified by human forecasters to take into account factors such as emerging flux or imminent collisions with other ARs. This subjective and largely ``climatological'' forecasting method has demonstrated only limited success in predicting SMEs during the 3-day forecast windows \citep{Sharpe&Murray:2017,Crown:2012}. For several decades, efforts have been made to automate the flare prediction process with computer-based classification and prediction systems in order to improve upon the current process. In recent years, these efforts have taken advantage of the rapid innovation in ``machine learning'' (ML) techniques developed for commercial image classification purposes. \cite{Qahwaji&Colak:2007} review some of the early attempts to employ machine learning to the solar flare prediction problem, and a recent series of papers compares the prediction skill of the manual method and several current automated models \citep{Barnes_survey,leka2019a,leka2019b,Park:2020}.  

The majority of automated flare prediction systems rely primarily on the properties of sunspot ARs derived from measurements of the one-dimensional line-of-sight (LOS or ``longitudinal'') component of the magnetic field in the solar photosphere. This is primarily because there are now almost 30 years of space-based full-disk photospheric magnetic field images (``magnetograms'') taken on cadences of order 10-min or less that provide continuous, consistent, and high quality data. However, the primary magnetic reconnection that triggers SMEs does not take place in the photosphere \citep[e.g.,][]{Simoes:2015}, and with few exceptions \citep[e.g.,][]{Sudol&Harvey:2005} photospheric magnetograms show no significant changes before and after SMEs. Ideally one would use magnetic field measurements in the upper solar atmosphere (the chromosphere and corona) to better predict eruption triggering, but there are not yet any reliably available, consistently high-quality, magnetic field measurements in the upper solar atmosphere. Thus solar physicists have concentrated on searching for physical properties derivable from photospheric magnetograms that correlate to a high probability of imminent eruption \citep[e.g.,][]{Schrijver:2016cs,Kusano:2020}.
Recently, the full-disk \emph{vector} (i.e., three-dimensional) magnetograms from NASA's Solar Dynamics Observatory \citep[SDO,][]{Chamberlin:2012} Helioseismic and Magnetic Imager \citep[HMI,][]{Scherrer2012} instrument have enabled the creation of a large database of derived AR magnetic field quantities called the SHARP parameters \citep{Bobra2014}. Since 2015, many ML solar flare prediction systems have used the SHARP parameters, or a subset of parameters, as the primary feature vector input to supervised learning architectures \citep[e.g.,][]{Bobra:2015fn,Bobra:2016,Florios2018,Chen2019,Deshmukh2020}. The SHARP AR image cutout and feature set has recently been expanded to include data from the predecessor instrument to SDO/HMI, the SOHO/Michelson Doppler Imager \citep[MDI;][]{Scherrer:1995}, to create the SMARPs dataset \citep{Bobra:2021}.

A common challenge for all ML-based solar flare prediction models is the relative dearth of large, space-weather important (defined as X-ray class M1 or above on the NOAA radio black-out scale\footnote{See \url{https://www.swpc.noaa.gov/noaa-scales-explanation} for definitions of the NOAA space weather scales.}) flares on which to train the models. Large SMEs and their associated large flares are relatively rare compared to the many smaller flares that occur in any given AR over the course of its evolution. Thus for any given sequence of AR magnetograms, there will be many more ``non-flare'' magnetograms, i.e, magnetograms that do not have an M1 or larger flare within the next \emph{k} hours, where \emph{k} is the forecasting window (typically 24, 48, or 72 hours), than there are ``flare'' labelled magnetograms. This fact, combined with the fact mentioned above that photospheric magnetograms show only minor changes before and after flares of any size implies that ML models will naturally train to predict no flaring, achieving high accuracy scores at the expense of sensitivity\footnote{See \cite{Jolliffe&Stephenson:2012} for formal definitions of these binary categorical forecasting metrics.}. This training set imbalance can be addressed in several ways. Balancing training sets by removing non-flare examples \citep[e.g.,][]{Chen2019} improves sensitivity, but such a model is likely to be difficult to optimize for operational space weather forecasting where the incoming real-time data stream is naturally extremely unbalanced. Other studies have addressed training set imbalance using oversampling of flare magnetograms during training \citep[e.g.,][]{Zheng2021} or data augmentation: creating artificial flare magnetograms using image processing techniques such as affine transformations of real flare magnetograms or employing Generative Adversarial Networks \citep[GANs;][]{Zheng2019}. Data augmentation has been successful in improving the training of ML image classification models \citep[e.g.,][]{wang:2017}, but our experiments with augmentation via affine transformation of flare magnetograms did not show improved skill over non-augmented training datasets (see Sec.~\ref{sec:results}). Another method of addressing training set imbalance that preserves the original dataset is to overweight the loss function used to train the network weights to penalize false negatives (missed flare detection) more severely than false positives. Models trained in this way can achieve high skill metrics in testing, but tend to overpredict flares, resulting in unacceptably high False Alarm Rates (FARs).

We have recently undertaken the development of ML models for prediction of solar flares based on Convolutional Neural Network (CNN) architectures that analyze spatial and, when used in recurrent architectures, temporal evolution of magnetic structure prior to flaring. Here we present a preliminary CNN model that analyzes SDO/HMI radial field (``$\mathrm{B_r}$'') magnetograms from the SHARP AR cutout series to produce a short-term (12-hour) probabilistic flare prediction. We achieve temporal correlation analysis by feeding the CNN several magnetograms in a temporal sequence as a single ``multi-layer'' input. This ``temporal stacking'' CNN input has been found to be more successful than the more traditional recurrent Long Short-Term Memory architecture models \citep{Hochreiter1997}. We use loss function weighting to compensate for training set imbalance and tune hyperparameters using a new ``scaled True Skill Statistic'' ($\text{TSS}_{\text{scaled}}$) metric to optimize the flare/no-flare threshold of the CNN model. We address the relatively high FAR by developing a hybrid architecture that employs an additional extremely randomized trees (ERT)
model. The ERT uses the flaring probability for a given magnetogram time series from the CNN stage as an additional feature added to derived features including the SHARPs parameters from the given set.


\section{Data}
\label{sec:data}

For this model, we use vector magnetogram image cut-outs observed by the Helioseismic
and Magnetic Imager (HMI) instrument on-board the Solar Dynamics Observatory
telescope \cite{Pesnell2012}. These cut-outs --- called as Spaceweather HMI Active Region Patch or \textit{SHARPs}
--- have been tracking active regions on the surface of the Sun visible to the SDO
since 2010 \citep{Bobra:2014dn}. For our dataset, we choose all magnetogram images, in the Cylindrical Equal Area (CEA) projection, across all recorded active regions from 2010 to 2017, at a cadence of 3 hours. This gives us a total of 157095 images.
Each image includes metadata that contains features associated with
the magnetogram including physics-based attributes extracted from the raw magnetic field
data and deemed important by solar physicists. 
X-ray irradiance data from the NOAA Geostationary Observational Environmental Satellite (GOES) provide the location, intensity, and the onset, peak and termination
times of recorded flares. Solar flares, based on the logarithm of the magnitude of their 1--8~\AA\ X-ray irradiance, are classified into 
five major categories --- A, B, C, M and X (in increasing order of magnitude). The first
three classes usually are considered as minor flares, while the remaining two are major 
flares, and therefore of higher importance. 
Combining the SHARPs metadata and the GOES flare data, we can determine if a magnetogram
produced a major flare (M- or X-class) in the next $k$ hours.  Since we are interested in short-term predictions that could generate solar flare warnings
we set $k = 12$ for this study. Each magnetogram image is labeled as 1 if it produced an
M/X flare within the following 12 hours, 0 otherwise. Since major flares are rare,
this labeling results in an extremely imbalanced dataset. 1561 ($\approx$1\%) of the total 
magnetograms are labeled positive (flaring), and the rest of the 99\% are labeled
negative. Such a highly imbalanced dataset poses a challenge for training ML models.

Existing CNN flare prediction models
\citep[e.g.,][]{Huang2018, Park2018, Zheng2019, Li2020, Abed2021, Zheng2021} use balanced
datasets for training and evaluating the models. The balanced datasets in these works 
are either generated by undersampling 
the majority class (lower intensity or no-flares) 
or oversampling the minority class (higher intensity flares)
for both the training and testing sets. 
As mentioned, data augmentation is used successfully in image classification research to balance datasets. We applied this technique to augment the minority
class of the training set by applying simple rotation and polarity 
swapping to generate new flaring magnetograms. This reduced the dataset imbalance to 1:10. However we found that our model showed no improvement in flare prediction skill as measured by, e.g., the TSS of the test set. As a result, we decided not to include data augmentation
in our experiments. To our knowledge the only ML flare prediction study to train on imbalanced data was \cite{Huang2018}, and we observe that their model suffers from a high false positive rate
similar to our CNN implementation described below. 

The magnetogram image dataset is not directly usable as-is with deep learning models. Each image cut-out 
is variable in size, whereas the convolutional neural network we model requires fixed input dimensions
across the entire dataset. There are multiple ways to transform all images to a standard size; in this
work, we choose to perform affine transformations, which is a linear transformation that preserves lines
and parallelism in an image, but not distances. We use the standard {\tt OpenCV} package to convert all SHARPS radial magnetograms to a $128\times128$ pixel format. We find that these are the smallest set of dimensions that require less memory and processing time without affecting the quality of predictions. 

To train and evaluate our architecture, we split our dataset into 70\% training, 10\% validation 
and 20\% testing sets. The splitting is based on the active region number, so that all images of
any given active region are present in the same set. The splitting 
is randomized 10 times using 10 random seeds; the model is trained,
tuned and tested one randomized dataset at a time, and the statistics of the
performance score reported across these 10 trials. With this 
arrangement, the total number of samples in the testing set 
is approximately 24000 samples, the positive samples varying from 
137--347 and negative samples between 22187--24532. The positive
and negative samples for the individual splits are shown in 
Table~\ref{tab:comparison_raw} in the appendix.

\subsection{Feature sets}
\label{sec:features}

We use the SDO HMI radial field ($\mathrm{B_r}$) magnetograms as input to a CNN model,
as well as a source to extract numerical subsets of features that are used to train
an extremely randomized trees (ERT) model, as described in Section~\ref{sec:model}. 
Here, we discuss the two types of numerical features extracted from the magnetogram
data. 

\subsubsection{Physics-based Features}

The first subset of features are the standard attributes of an active region cut-out available in the metadata
of the SDO/HMI SHARPs dataset. These attributes, such as the area, total magnetic
flux, magnetic shear, total vertical current, current helicity, etc., are predominantly derived from the 
spatial and/or extensive properties of the vector magnetic field in a given magnetogram image. A complete 
list of these features is available in Table~\ref{tab:sharps_features}. 

\begin{table}
\begin{center}
\begin{tabular}{| l | l | l |}
\hline
Acronym & Description & Units\\
\hline
LAT\_FWT & Latitude of the flux-weighted center of active pixels & degrees \\
LON\_FWT & Longitude of the flux-weighted center of active pixels & degrees \\
AREA\_ACR & Line-of-sight field active pixel area & micro-hemispheres \\
USFLUX &  Total unsigned flux & $Mx$ \\
MEANGAM &  Mean inclination angle, gamma & $degrees$ \\
MEANGBT &  Mean value of the total field gradient & $G/Mm$ \\
MEANGBZ &  Mean value of the vertical field gradient & $G/Mm$ \\
MEANGBH &  Mean value of the horizontal field gradient & $G/Mm$ \\
MEANJZD &  Mean vertical current density & $mA/m^2$ \\
TOTUSJZ &  Total unsigned vertical current & $A$ \\
MEANALP &  Total twist parameter, alpha & $1/Mm$ \\
MEANJZH &  Mean current helicity & $G^2/m$ \\
TOTUSJH &  Total unsigned current helicity & $G^2/m$ \\
ABSNJZH &  Absolute value of the net current helicity&  $G^2/m$ \\
SAVNCPP &  Sum of the absolute value of the net currents per polarity & $A$ \\
MEANPOT &  Mean photospheric excess magnetic energy density & $ergs/cm^3$ \\
TOTPOT &  Total photospheric magnetic energy density & $ergs/cm^3$ \\
MEANSHR &  Mean shear angle (measured using $B_{total}$) & $degrees$ \\
SHRGT45 &  Percentage of pixels with a mean shear angle greater than 45 degrees & $percent$ \\
R\_VALUE & Sum of flux near polarity inversion line & $G$ \\
NACR & The number of strong LOS magnetic field pixels in the patch & N/A \\
SIZE\_ACR & Projected area of active pixels on image & micro-hemispheres \\
SIZE & Projected area of patch on image & micro-hemispheres \\
\hline
\end{tabular}
\end{center}
\caption{The SHARPs feature set, as available in the metadata of the SDO HMI 
  dataset. Abbreviations: $Mx$ is Maxwells, $G$ is Gauss, $Mm$ is Megameters, 
  and $A$ is Amperes. From \cite{Bobra:2014dn}.
  }
\label{tab:sharps_features}
\end{table}

\subsubsection{Shape-based Features}

To complement the physics-based features, we also include some shape-based features
extracted using topological data analysis (TDA), as proposed in \cite{Deshmukh2020}.
TDA is an approach to characterize the shape of data in terms of its homology, i.e.
by counting the $j-$dimensional holes of an object. The counts of these $j$-dimensional
holes are defined as Betti numbers $\beta = \lbrace\beta_0, \beta_1, \beta_2, ...\beta_{d-1}\rbrace$,
where $d$ is the dimension of the manifold that the data lies in. 
$\beta_0$ counts the number of connected components in an object, $\beta_1$ the number of circular 
2-dimensional loops, $\beta_2$ the total number of 3-dimensional voids, and so on. Since
we are dealing with a 2-dimensional image for extracting the features, our Betti numbers
are restricted to $\lbrace\beta_0, \beta_1\rbrace$. As in \cite{Deshmukh2020}, we 
choose $\beta_1$ for our topology-based feature set.

On an image, TDA counts holes by first performing sub-level thresholding, i.e. keeping
magnetic flux pixels below a chosen threshold and discarding the rest. The selected pixels
connect to each other forming connected components ($\beta_0$) and loops with empty space
between them ($\beta_1$). Repeating this process for 7 thresholds on the 
positive and negative flux structures on a magnetogram separately, we obtain the $\beta_1$ 
counts for each of the thresholds. We choose equally spaced magnitudes of magnetic flux
thresholds for the positive and negative fluxes,
\[
thresholds = \lbrace20G,\ 420G,\ 820G,\ 1220G,\ 1620G,\ 2020G,\ 2420G\rbrace.
\]
This gives us a total of 14 TDA-based features, denoted by {\tt flux\_pos\_t} and
{\tt flux\_neg\_t}, where $t \in thresholds$. 

\section{A two-stage machine learning pipeline}
\label{sec:model}

In this study, our primary machine learning model is a CNN. 
CNNs are an effective tool for extracting patterns from 
images which can subsequently be used in training the model to automatically classify the image content. This is advantageous as a way to avoid manual feature engineering of the dataset, or alternatively 
as a way to complement these manually engineered features. Because the pattern extraction is statistical in nature and not easily attributed to any particular heuristic, this process is sometimes referred to as ``deep learning.'' Here we aim to combine the predictive power of the manually
engineered features (SHARPs and topological) of the magnetograms with the
features automatically extracted by a CNN model. 

To do so, we propose a two-stage model architecture. The first stage implements
a CNN architecture that is trained on the magnetogram images directly and 
classifies a flaring probability as the output. This flaring probability is 
then used as a feature along with the manually engineered features to train an extremely
randomized trees (ERT) architecture. We choose this architecture on account of its design 
simplicity in being able to separate the prediction capabilities of the CNN features
and the engineered features. An additional benefit of using the ERT architecture 
is its ability to rank the relevance of various features in terms 
of predicting flares. We discuss the two stages below. 

\subsection{Stage I: Convolutional Neural Network}
 
The first stage of the model is a CNN adapted from the VGG-16 model - a deep CNN with 13 2-D convolution layers and 4 dense layers designed to classify images into 1000 pre-defined categories \citep{Simonyan2014}. The input sample to our model is not a single magnetogram image, but a temporal stack of 4 consecutive magnetograms separated by a cadence of 3 hours. The input layer and the output layer of the VGG-16 model are modified to have 4 channels instead of 3, and two output nodes instead of 1000, respectively. The two output nodes respectively produce the probability of flaring and non-flaring in the next $k$ hours for a given input sample, where $k$ is the forecast window, which can be distinct from the 12-hour temporal stack we provide as input to the model. For this study however we set $k = 12$ to match the forecast window with the temporal span of the input data. Probabilistic output is desirable from a forecasting point of view. However, for comparison
to most other automated solar flare prediction models that produce categorical flare/no-flare event predictions, we convert the probabilistic output into a categorical output by defining an optimal flare event threshold. Optimizing the threshold is accomplished using validation data, as described in Section~\ref{sec:tuning}. Some methods choose an arbitrary threshold of 0.5 to generate the categorical
prediction \citep{leka2019a} while others use an automatically determined threshold based on optimizing the Receiver Operating Characteristic \cite[ROC,][]{Jolliffe&Stephenson:2012}. 
Interestingly, this coincides with the threshold that maximizes the TSS
score. 

Four variations on the model architecture and the input data format were investigated: 
\begin{enumerate}
    \item $C_{1}$: Using an input stack consisting of the [$B_r, B_\phi, B_\theta$] components of the vector magnetogram and setting the number of input channels of the VGG-16 model to 3. 

    \item $C_{2}$ Using a single component $B_r$ at a single time as the input data per sample, setting the number of input 
    channels of the VGG-16 model to 1. 
    
    \item $C_{3}$: Using the temporal stacked configuration [$B_{r,t}, B_{r,t-3}, B_{r,t-6}, B_{r,t-9}$] as 
    the input data per sample (as described above), setting the number of input channels of the VGG-16 model to 1. Each component in the temporal stack is operated on by the convolutions and dense layers individually to generate 4 feature representations. An LSTM layer is introduced before the output layer to process the sequence of the 4 representations. 
    
    \item $C_{4}$: Using the temporal stacked configuration [$B_{r,t}, B_{r,t-3}, B_{r,t-6}, B_{r,t-9}$] as 
    the input data per sample (as described above), setting the number of input channels of the VGG-16 model to 4. All components of the temporal stack are treated as individual channels in the input layer. Such a setup leads to a single feature representation at the final layer that is acted upon by the softmax function, as opposed to the four representations generated from the temporal stack in $C_3$.
    
\end{enumerate}
All four configurations were modeled in {\tt Pytorch}.  We use a weighted 
focal loss function ($\alpha = \frac{1}{11}$, $\gamma = 2$) \citep{Lin2017} with an additional
$L_2$ regularization weight decay factor $\beta = 0.001$. The weights of the models are updated using
an {\tt Adagrad} optimizer \citep{duchi2011}, using an initial learning rate of $0.0001$ and a 
batch size of $64$. A cosine annealing learning rate scheduler is used for adjusting the learning rate through the training.
The evaluation of each of these configurations (after hyperparameter tuning discussed 
below), is presented in the appendix in Table~\ref{tab:cnn_only}. Summarizing the results, we find that the 
$B_{r}$-only configuration $C_{2}$ performs slightly worse than the vector magnetogram configuration 
$C_{1}$. However, the temporal stacking configuration $C_{4}$ performs very similar to $C_{1}$.
What this tells us is that from the perspective of the CNN, the $B_r$ channel is sufficient
for predicting flares and the other two components are superfluous. Additionally, 
comparing the two temporal stacking configurations, the LSTM model in $C_{3}$ does
worse than using the temporal stack as channels as in $C_{4}$.

\subsection{Stage II: Extremely Randomized Trees (ERT) model}

While there have been CNN implementations proposed in recent years, the novelty 
in our approach is the combined use of the features extracted through convolutions
together with engineered features based on the physics and the shape of the
magnetogram. We generate a feature set that concatenates the output of the VGG-16 model 
(the probability of a flare) with these engineered features. These features are extracted from two
main sources. The first is a set of physics-based features --- called \textit{SHARPs}
--- that are available in the SHARPs HMI data set as metadata \citep{Bobra2014}. 
The second source is a set of features extracted using topological data analysis (TDA)
\citep{Zomorodian11},
as applied to sunspot magnetograms in \citet{Deshmukh2020, Deshmukh2021}. 
Concatenating the VGG-16 probability output, 20 SHARPs features and 14 TDA-based 
features, we have a complete feature set of 35 features which combines information
from deep learning-based and feature engineered approaches. 

\begin{figure}
    \centering
    \includegraphics[width=\textwidth]{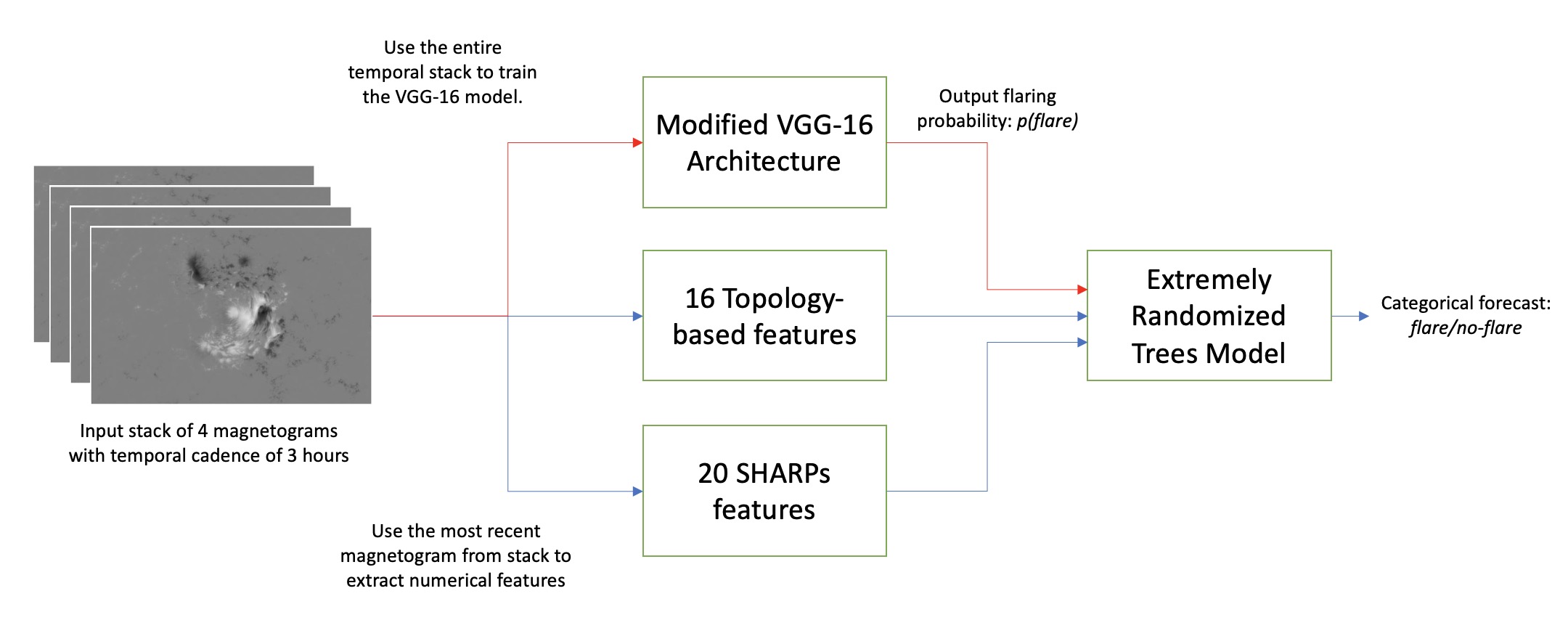}
    \caption{Our two-stage model for solar flare prediction. The input is a temporal stack of B$_r$ magnetograms from SDO/HMI which is both fed to a custom CNN model and analyzed for feature vectors. The CNN model outputs the probability of flaring with the 12-hour forecast window and this probability is combined with the feature vectors to create a single feature vector input to the ERT model. The output of the ERT model is a binary event prediction.}
    \label{fig:model}
\end{figure}

We use this feature set to train an extremely randomized trees (ERT) model 
\citep{Geurts2006} in the second stage. An ERT is a tree-based model built
as a hierarchical structure of nodes that successively perform the operation
of separating the dataset into two classes. The entire dataset is ``fed" to the
root of this tree and undergoes a sequence of splitting operations at the 
intermediate nodes. At each node, the incoming dataset is
separated into two subsets --- termed ``left" and ``right" --- based on a feature 
thresholding criterion. That is, a random subset
of $m$ features is chosen from the entire candidate feature set, together 
with $m$ random splits (one for each feature). The quality of each split at node $n$ is $s_n$, 
determined by computing the reduction in some ``impurity" metric
of the dataset given by --- 
\begin{equation}
\Delta i(s_n, n) = i(n) - p_L \times i(n_L) - p_R \times i(n_R).
\end{equation}
Here, the impurity function $i(n)$ quantifies the degree of class intermixing for a given 
dataset input into node $n$. Correspondingly, 
the impurities for the left and the right subsets from the split are denoted by $i(n_L)$ 
and $i(n_R)$ respectively. $p_L$ = $N_{n_L} / N_n$
and $p_R = N_{n_R} / N_t$ represent the proportions of the dataset arriving at node $n$ of size
$N_n$ split into the left (size $N_{n_L}$) subset and right subset (size $N_{n_R}$) respectively. 
Of the $m$ splits, the one that maximizes $\Delta i(s_n, n)$ is chosen. For the definition of 
impurity, we choose the standard Gini impurity index, as described in \cite{raileanu2004}.
Whether the two subsets are further subject to splitting at the next level is determined by an
important hyperparameter in the training process known as the {\tt min\_impurity\_decrease\_index}, 
denoted by $\Delta i_{min}$. A dataset at any point in the tree 
is split further using an additional node if,
\[
\Delta i(s_t, t) \ge \Delta i_{min}.
\]
In the context of this problem, $\Delta i_{min}$ determines how the model balances between the 
true positive rate (TPR) and
false positive rate (FPR, also called the False Alarm Rate). A low value of $\Delta i_{min}$ results in low FPR but a low TPR as well, whereas a high $\Delta i_{min}$ raises the TPR at the
cost of increased FPR as well. Just as with the threshold in the CNN stage, 
we tune this hyperparameter using a validation set (discussed in Section~\ref{sec:tuning}). 
The two-stage model is summarized in Fig.~\ref{fig:model}.

\subsection{Metrics}
\label{sec:metrics}

With a categorical forecast (flare/no-flare), we can compute the standard confusion
matrix on the testing set predictions: true positives (TP), false positives (FP),
true negatives (TN) and false negatives (FN). Using the entries of the confusion
matrix, we study various metrics defined in Table~\ref{tab:metrics}. Most of these
metrics are standard to flare prediction literature. A popular one among these
is the true skill statistic score (TSS), equal to the difference TPR - FPR.
TSS provides some utility
to this problem because it is insensitive to dataset imbalance, and is a better
indicator of the model performance than the standard accuracy \citep{Barnes:2016bu, Bobra:2015fn}. However, optimizing the TSS score often leads to an overforecasting model;
models optimized on TSS tend to improve TPR at the cost of also slightly 
increasing the FPR. A slight increase in FPR can lead to a significant increase in
the absolute false positives FP, since the number of negative samples is huge, 
thereby impacting other metrics like precision or F1, that are sensitive to FP.

\begin{table}[htbp]
    \centering
    \begin{tabular}{c | c}
    Metric & Formula  \\
    \hline
    \hline
    Recall (TPR) & $\frac{\text{TP}}{\text{TP + FN}}$ \\
    False Positive/Alarm Rate (FPR) & $\frac{\text{FP}}{\text{FP + TN}}$ \\
    Accuracy & $\frac{\text{TP + TN}}{\text{TP + TN + FP + FN}}$ \\
    Precision & $\frac{\text{TP}}{\text{TP + FP}}$ \\
    True Skill Statistic (TSS) & $\frac{\text{TP}}{\text{TP + FN}} - \frac{\text{FP}}{\text{FP + TN}}$ \\
    Heidke Skill Score (HSS) & $\frac{\text{TP} \times \text{TN} - \text{FP} \times \text{FN}}{\text{(TP + FP)(FP + TN) + (TP + FN)(FN + TN)}}$\\
    \hline
    \end{tabular}
    \caption{Metrics used for evaluating the binary forecasting models.}
    \label{tab:metrics}
\end{table}

To address this problem, we define a new metric for model optimization, $\text{TSS}_{\text{scaled}}$, given by
\begin{equation}
    \text{TSS}_{\text{scaled}} = \text{TPR} - \frac{\text{TPR}_{\text{max}}}{\text{FPR}_{\text{max}}}\text{FPR},
\end{equation}
where $\text{TPR}_{\text{max}}$ and $\text{FPR}_{\text{max}}$ are the maximum values of the two metrics determined over the range of model hyperparameters. The scaling factor
$\frac{\text{TPR}_{\text{max}}}{\text{FPR}_{\text{max}}}$ applied to the FPR term
is a quantity greater than 1 for our hyperparameter choices that penalizes increase
in false positives more than the TSS score.   
We therefore choose $\text{TSS}_{\text{scaled}}$ for selecting the model 
hyperparameters, which, as we will show in the next section provides a better
balance between TPR and FPR. 

\subsection{Hyperparameter Tuning}
\label{sec:tuning}

\begin{figure}[htbp]
	\centering
	\gridline{\fig{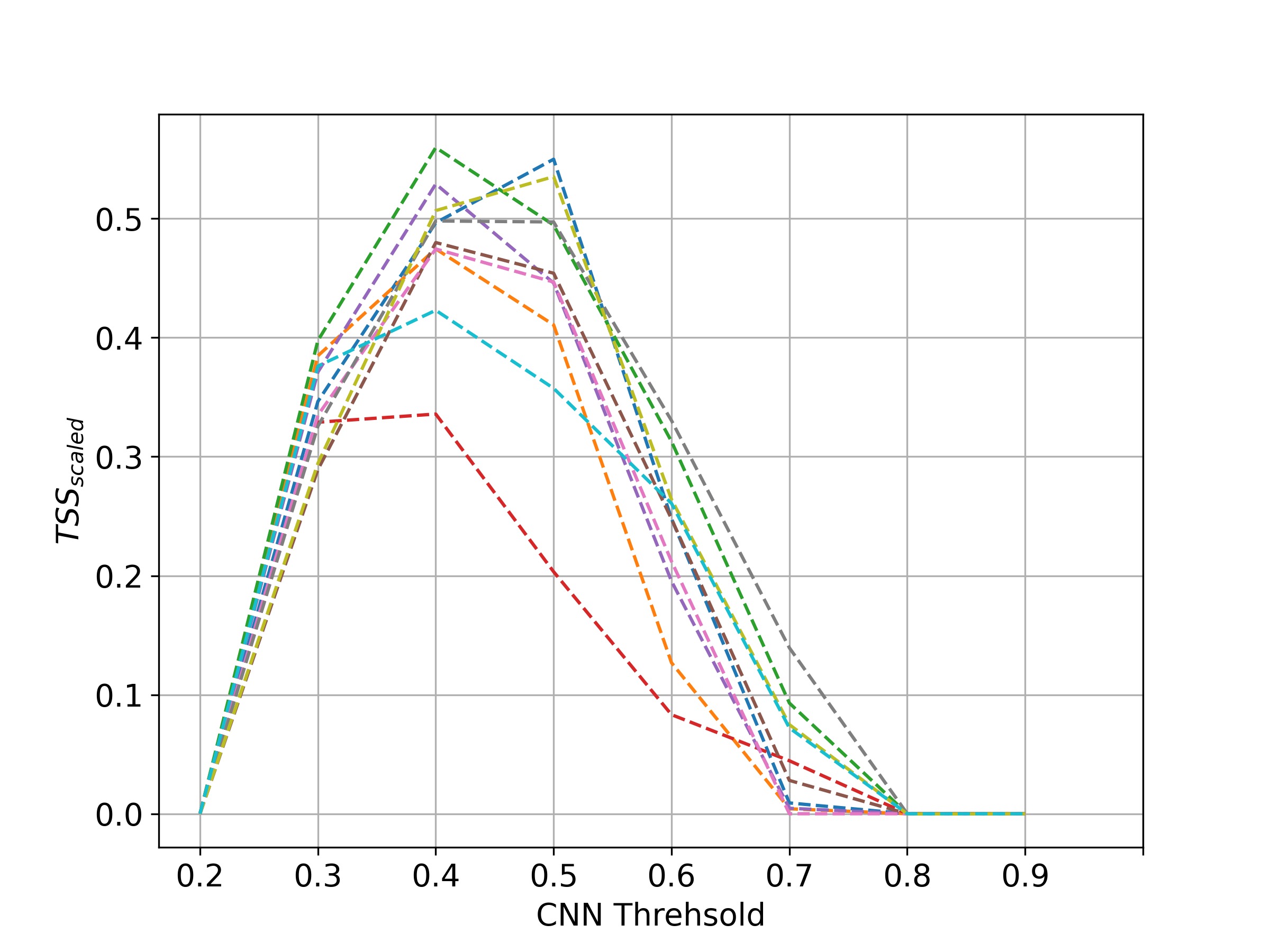}{0.5\textwidth}{(a) Stage-1 CNN-only}
		\fig{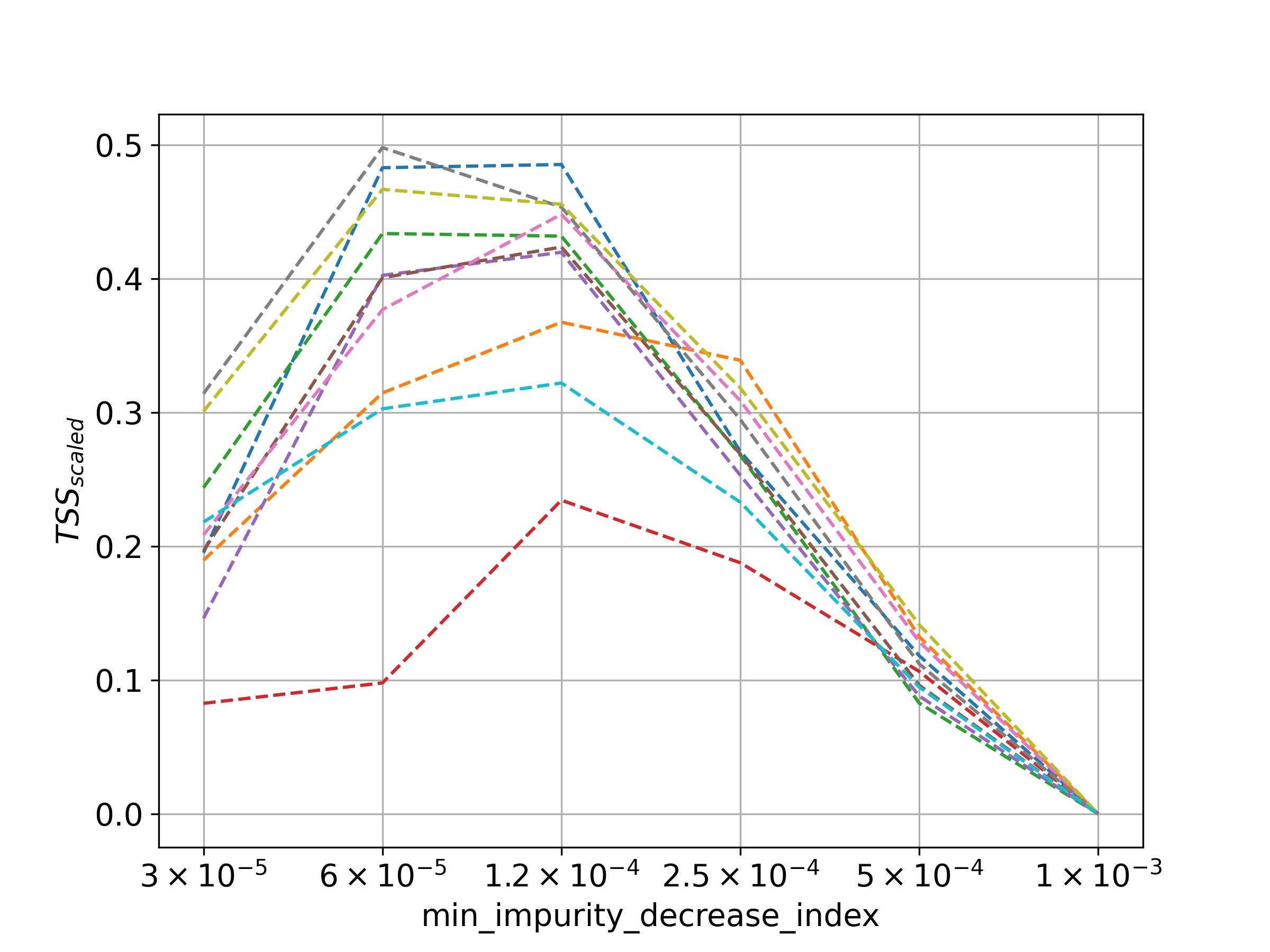}{0.5\textwidth}{(b) Stage-2 CNN+ERT}
	}
    \caption{Hyperparameter tuning across multiple seeds for both stages of the hybrid 
    flare prediction model. In both stages, hyperparameters that optimize the $\text{TSS}_{\text{scaled}}$
    metric are determined.}
    \label{fig:validation_results}	
\end{figure}

Hyperparameter tuning is essential for optimizing machine leaning model performance.
For both models, we identify the hyperparameter that significantly
affects the TPR-FPR balance. In case of the first stage CNN-only
model, it is the threshold for converting probabilistic to categorical
forecast. In case of the second stage ERT model, it is the 
{\tt min\_impurity\_decrease\_index} parameter. 

The process for choosing the optimal hyperparameters is performed individually on the CNN and ERT. 
The first step is to choose a set of suitable hyperparameter values to sample from in each case. 
The CNN stage is trained only once, determining the $\text{TSS}_{\text{scaled}}$ score by 
simply choosing different thresholds 
on the validation set. The ERT model is trained with all values of the chosen 
hyperparameter set.The setting that maximizes
the chosen metric on the validation set --- in this case, $\text{TSS}_{\text{scaled}}$
--- is then determined.

This process is applied to both stages across the 10 randomly selected train-validation-testing
set combinations. For the CNN-only stage, the tuning is performed over a 
range of eight even spaced thresholds in the interval $[0.2, 0.9]$. In the second
stage, six values of {\tt min\_impurity\_decrease\_index} are chosen for tuning --- 
$[3\times10^{-5}, 6\times10^{-5}, 1.2\times10^{-4}, 2.5\times10^{-4},
5\times10^{-4}, 1\times10^{-3}]$
The results are shown in Fig.~\ref{fig:validation_results}. 
It is clear that
for the majority of the dataset combinations, a single hyperparameter maximizes
$\text{TSS}_{\text{scaled}}$. For stage 1, this is a threshold of 0.4, for the second,
{\tt min\_impurity\_decrease\_index}$ = 0.00012$. Performing a similar optimization
study for the TSS score instead yields an optimal threshold of $0.3$ for stage 1
and an optimal {\tt min\_impurity\_decrease\_index}$ = 0.001$. Note that if the
CNN+ERT model were tuned with this latter value of 0.001, all 10 trials would have converged onto a $\text{TSS}_{\text{scaled}}$ score of zero. The TPR and FPR statistics on 
the validation set across the 10 trials for each of the these hyperparameter choices are shown in 
Table~\ref{tab:tpr_fpr_ranges}. It can be seen that optimizing over $\text{TSS}_{\text{scaled}}$
over optimizing TSS on average reduces FPR by a factor of approximately 2-4 
while only reducing the TPR by a factor of approximately 1.2-1.4. The $\text{TSS}_{\text{scaled}}$
optimized hyperparameters offer a more favorable result in terms of the TPR-FPR balance.
For our models, we therefore use the hyperparameters that optimize $\text{TSS}_{\text{scaled}}$ on the
validation set. 

\begin{table}[htbp]
    \centering
    \begin{tabular}{c|c|c|c|c}
         Model stage & Optimized metric & Optimal hyperparameter & TPR & FPR \\
        \hline
        \hline
         CNN-only & TSS & threshold = 0.3 & $0.90 \pm 0.05$ & $0.13 \pm 0.02$ \\
         CNN-only & $\text{TSS}_{\text{scaled}}$ &  threshold = 0.4 & $0.75 \pm 0.09$ & $0.06 \pm 0.01$ \\
        \hline
         CNN+ERT & TSS & {\tt min\_impurity\_decrease\_index} = 0.001 & $0.90 \pm 0.05$ & $0.12 \pm 0.02$ \\
         CNN+ERT & $\text{TSS}_{\text{scaled}}$ & {\tt min\_impurity\_decrease\_index} = 0.00012 & $0.65 \pm 0.11$ & $0.03 \pm 0.01$ \\
         
    \end{tabular}
    \caption{The mean and standard deviation values of TPR and FPR on the validation set ($10\%$ of the full dataset or $\approx 15,000$ samples)
    for the two stage models using TSS and $\text{TSS}_{\text{scaled}}$ metrics for optimizing hyperparameters. 
    The statistics are generated over 10 trials with 10 different random dataset splits.}
    \label{tab:tpr_fpr_ranges}
\end{table}
 
\section{Results}
\label{sec:results}
 
\begin{figure}[htbp]
	\centering
	\gridline{
		\fig{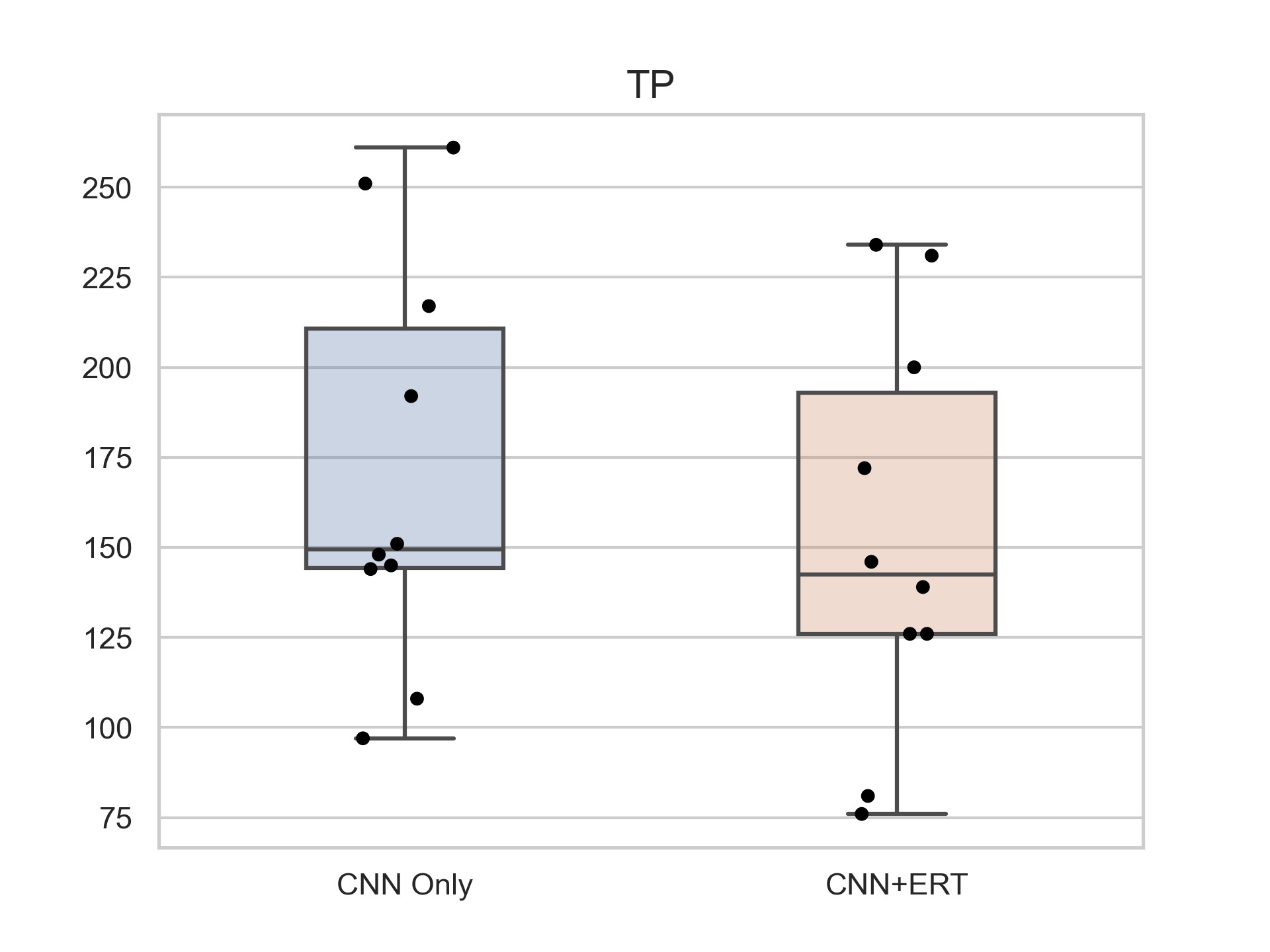}{0.4\textwidth}{(a) TP}
		\fig{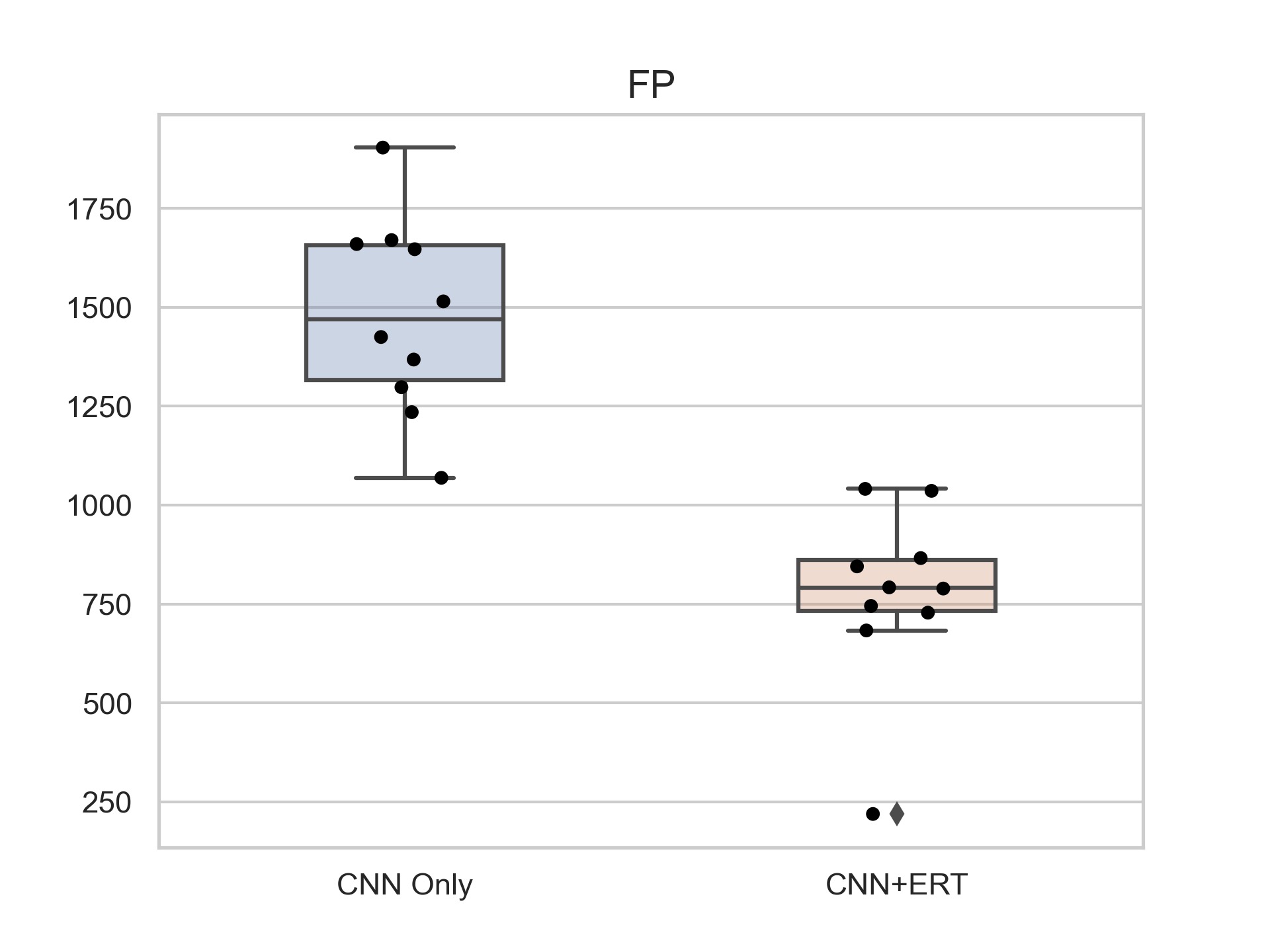}{0.4\textwidth}{(b) FP}
	}
	\gridline{
		\fig{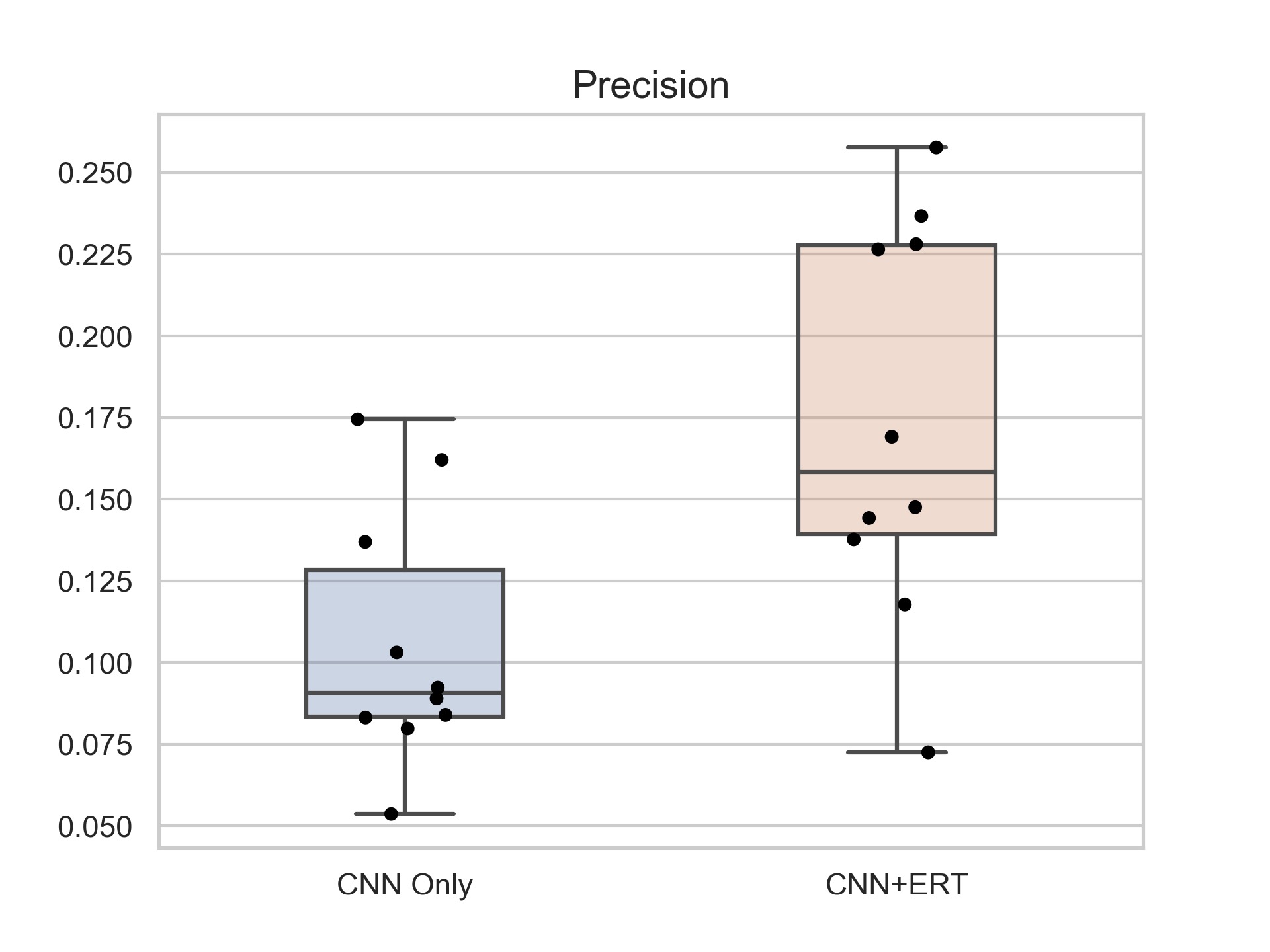}{0.4\textwidth}{(c) Precision}
		\fig{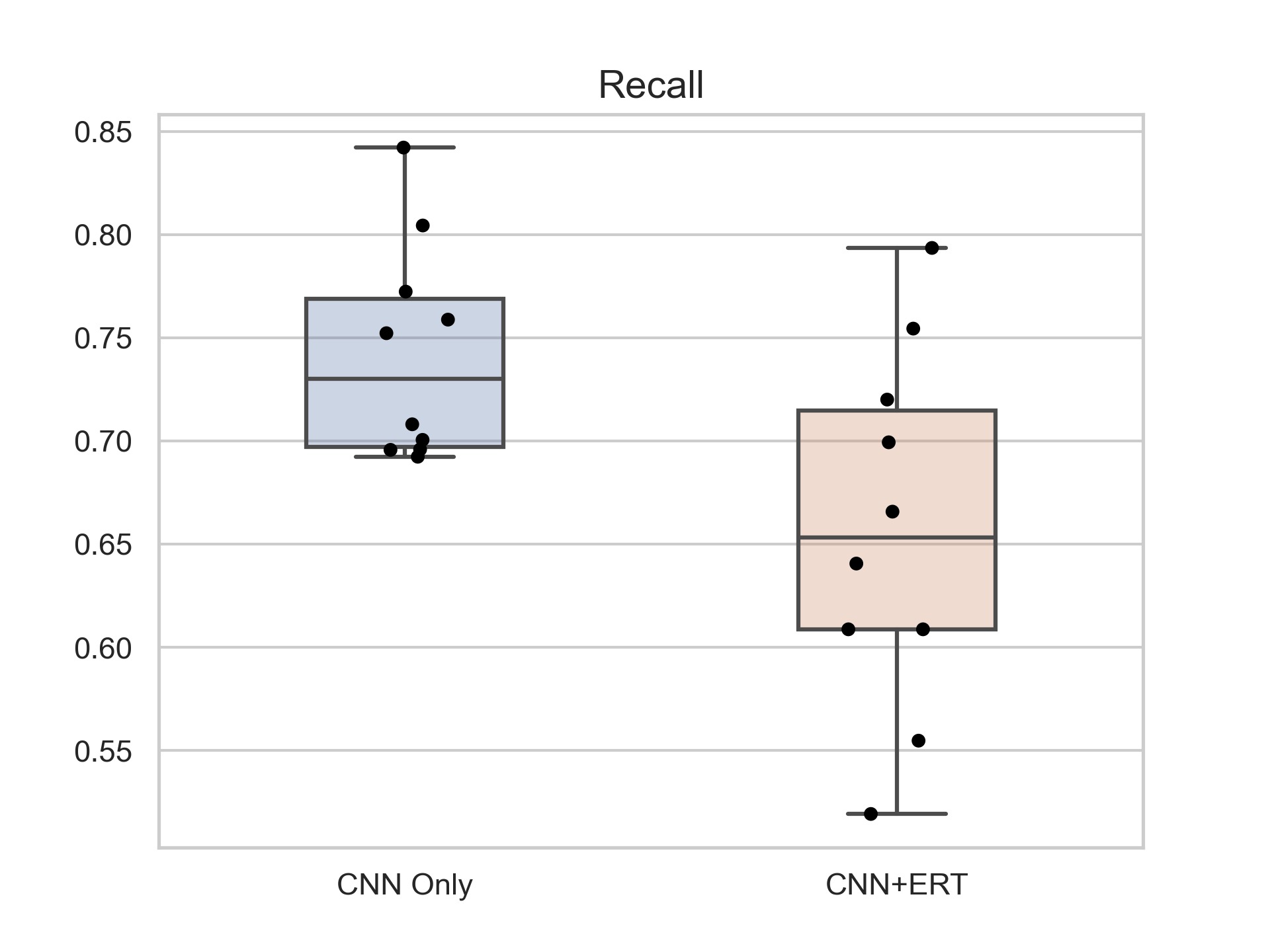}{0.4\textwidth}{(d) Recall}
	}
	\gridline{
		\fig{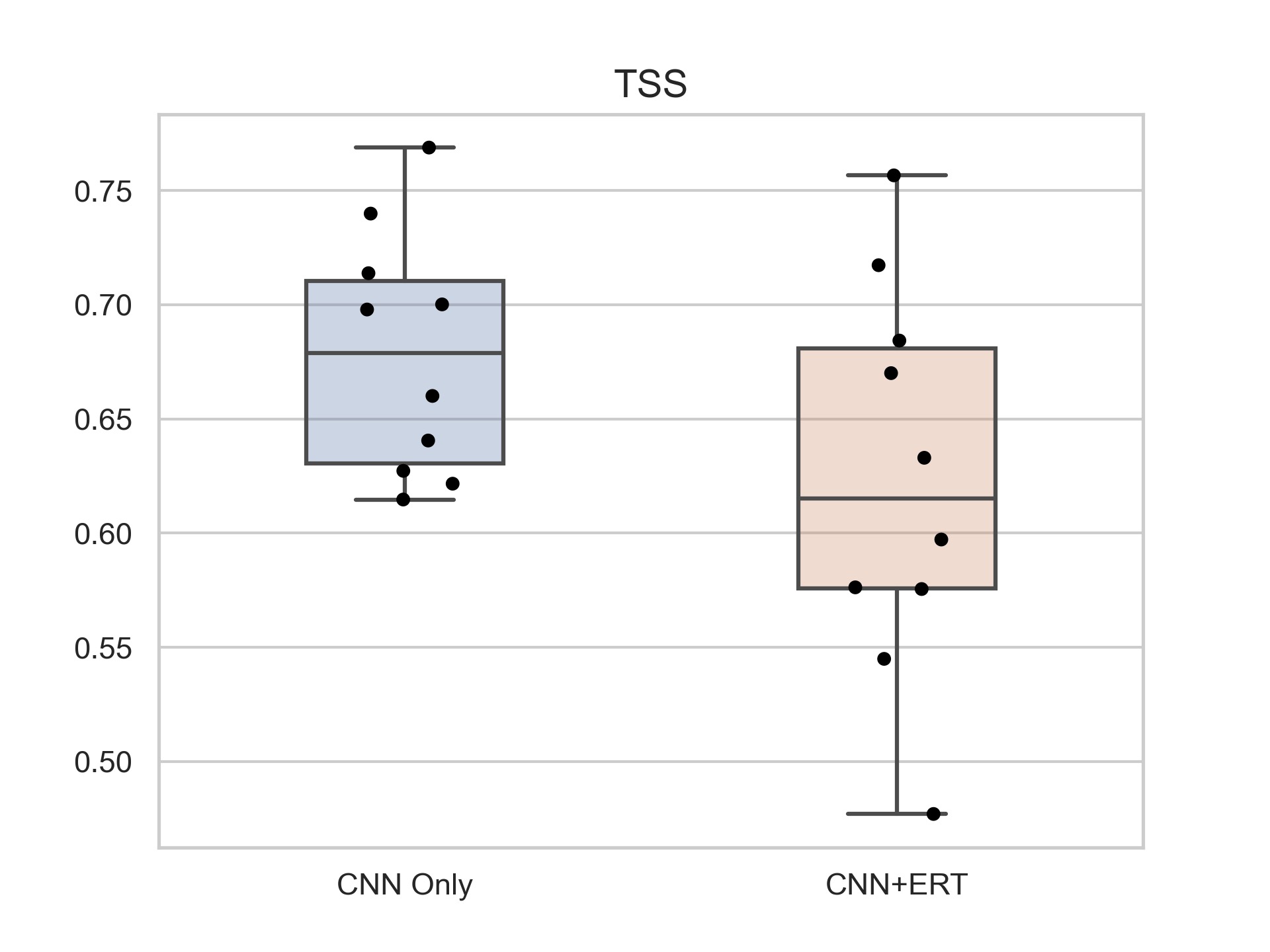}{0.4\textwidth}{(e) TSS}
		\fig{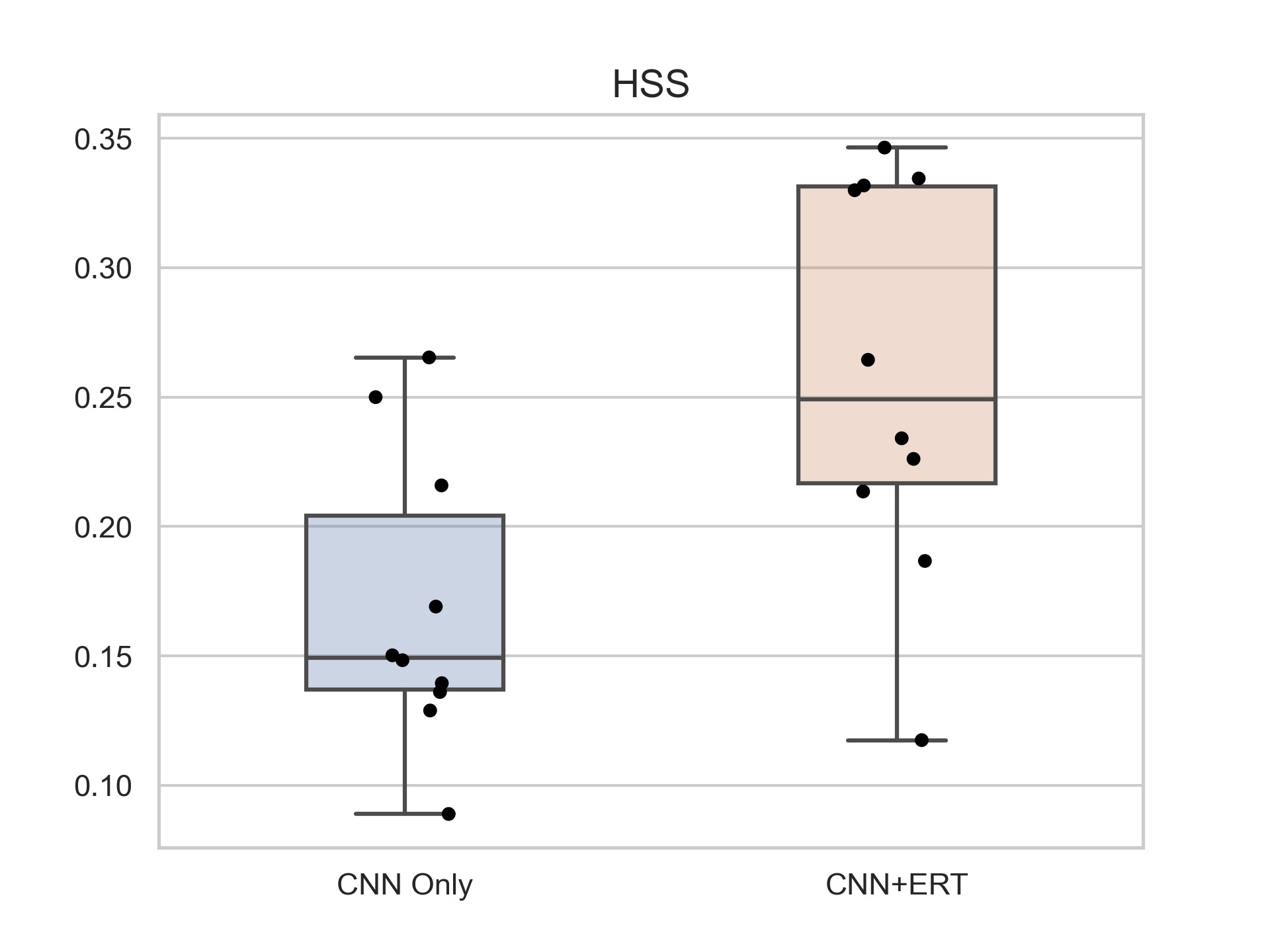}{0.4\textwidth}{(f) HSS}
	}
    \caption{Performance comparison between CNN-Only and CNN+ERT models across six different metrics on the testing set. For each metric boxplot, the 10 dots shown represent the individual score of each of the 10 dataset splits.}
    \label{fig:model_perf_all_metrics}
\end{figure}

For the 10 dataset splits, we separately determine the confusion matrix on the test set 
for each of the two stages, and calculate the metrics discussed in Table~\ref{tab:metrics}.
The optimal hyperparameters for each stage, as derived in Section~\ref{sec:tuning}, are used. 
We then evaluate the change in these various metrics between stages, i.e. using only the CNN and then appending the ERT to it. The raw values of six metrics is shown in Fig.~\ref{fig:model_perf_all_metrics}
in the form of box plots.  Note also that
TP and FP are presented in this plot instead of TPR and FPR. Looking at these two
metrics in box plots Fig.~\ref{fig:model_perf_all_metrics}(a) and (b), we observe that TP is slightly decreased with the use of the ERT architecture. On the
other hand the FP values decrease significantly, thus reducing the over-forecasting nature
of the model. It should also be observed that the FP box plots for the CNN-Only and CNN+ERT
architecture are non-overlapping, demonstrating that the improvement is significant. 
The changes in TP and FP scores impact other metrics both positively and negatively.
For example, the precision and the HSS score in Fig.~\ref{fig:model_perf_all_metrics}(c)
and (f) respectively are overall better for the CNN+ERT architecture, whereas the recall 
and TSS are overall slightly worse due to the dominance of TP in calculating these metrics. (Fig.~\ref{fig:model_perf_all_metrics}(d) and (e)).

 \begin{table}[htbp]
     \centering
     \begin{tabular}{c | c}
        Metric & \% average change in metric between stages \\ 
        \hline
        \hline
        Recall (TPR) & -12 $\pm$ 6.9 \\
        False Positive/Alarm Rate (FPR) & -48 $\pm$ 12.4 \\
        Accuracy & 3 $\pm$ 0.7 \\
        Precision & 69 $\pm$ 16.7 \\
        TSS & -8 $\pm$ 7.0 \\
        HSS & 56 $\pm$ 35.7 \\
        \hline
     \end{tabular}
     \caption{Percent change in metrics of using the 2-stage model (CNN+ERT) over using a 
     single stage CNN-only model, along with the standard deviation, summarized over 10 
     dataset experiments.}
     \label{tbl:perf}
 \end{table}
 
Table~\ref{tbl:perf} shows the average percentage improvement across all the dataset splits,
which summarizes the results in Fig.~\ref{fig:model_perf_all_metrics}. 
The true positive rate is decreased (on average) by 12\%, while the 
false positive rate improves by 48\%. This impacts the derived metrics in different ways. 
For example, the more popular TSS metric is decreased by an average of 8\%, and similarly the 
recall is decreased by an average of 12\%. On the other hand, we see very large improvements in precision ($\approx$69\%) and HSS ($\approx$56\%).
Thus, our two-stage model provides a prediction that can be more reliably incorporated into solar flare forecasting processes. We stress that operational flare forecasts are never dependent on a single model -- they always incorporate multiple factors including climatological predictions, model predictions, and changing conditions evaluated in real time by forecasters.

\subsection{Feature Ranking}
\begin{figure}
    \centering
    \includegraphics[width=1.1\textwidth]{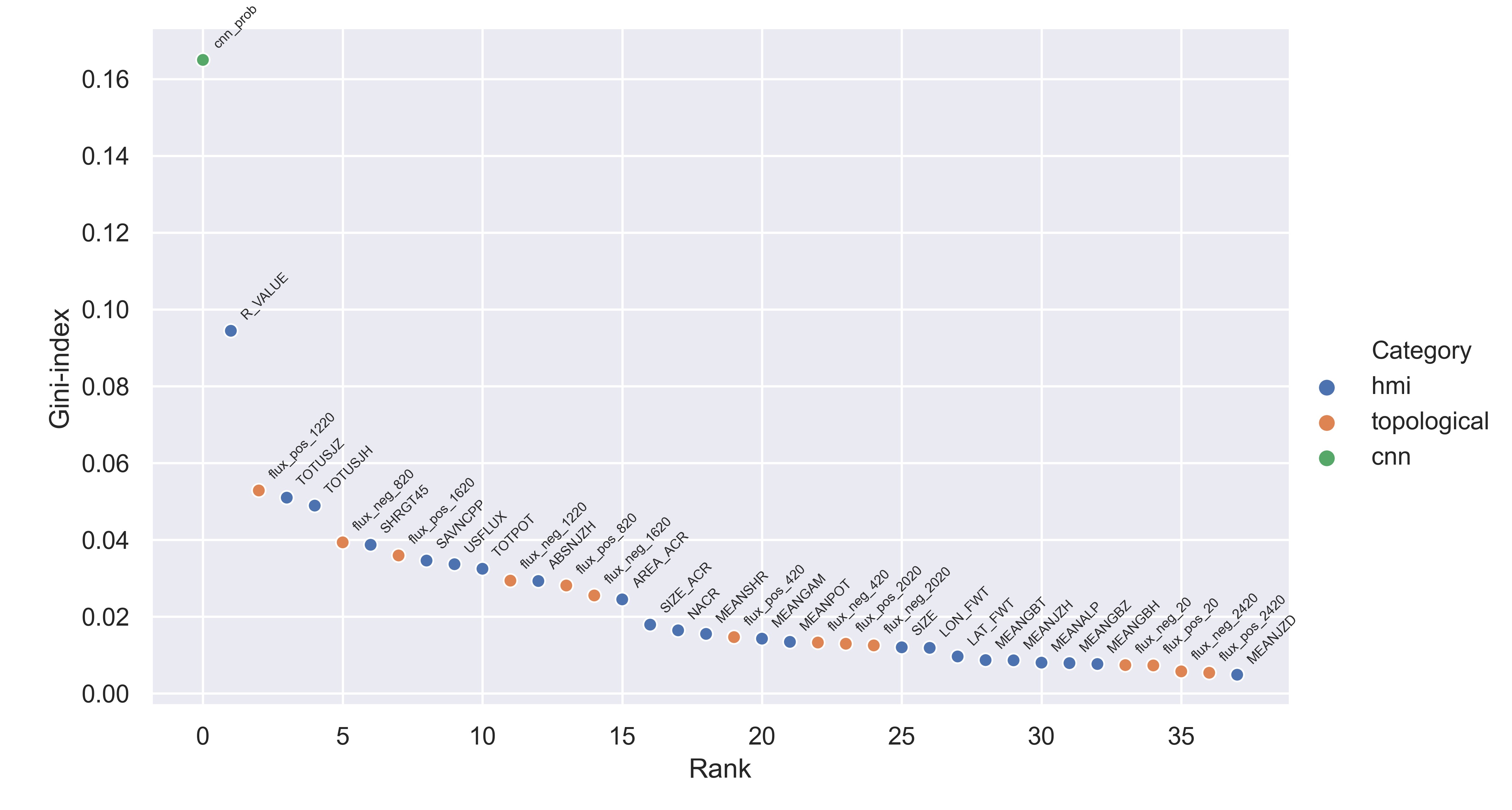}
    \caption{Feature ranking using the Gini impurity index from the ERT model.}
    \label{fig:feature_ranking}
\end{figure}

It is useful to know which features in our feature set play an important role in 
the model prediction. We use the Gini impurity index extracted from the ERT model to determine
how much each feature is successful in separating the positive and negative
labels across all the nodes of the tree. The relative rankings are shown in 
Fig.~\ref{fig:feature_ranking}. While there are other ways for performing
multivariate feature ranking, e.g. the linear discriminant analysis from 
\cite{Leka2007}, extending our ERT prediction model to also perform feature ranking makes sense. 


Fig.~\ref{fig:feature_ranking} shows that the CNN output probability {\tt cnn\_prob} ranks highest amongst all the features. 
In the top 10 features there is a mix of topological and physics-based features.
The {\tt R\_VALUE} feature from the HMI feature-set ranks highly (second only to the CNN probability), since this corresponds to the magnetic flux from the
neutral line where magnetic reconnection (and thus SMEs) occur.
With regard to the topological feature set, the $\beta_1$ counts in the range of 800G to 1600G are 
shown to be important.

 \section{Conclusions}
 \label{sec:conclusion}

Extreme dataset imbalance is a significant challenge for machine learning-based solar flare prediction. To address this problem, various methodologies have been adopted in previous literature. 
A common approach is to balance the dataset, either through oversampling the minority class or 
undersampling the majority class. Some approaches weight the loss function of the model that 
penalizes mispredictions on the minority class more in relation to the majority class. Finally, 
many methods optimize models to maximize metrics such as the True Skill Statistic (TSS). This study presents a systematic evaluation of some of these approaches, uncovering their limitations and proposing modeling and
evaluation strategies for overcoming them. To that end, we have proposed a two-staged machine learning 
model for predicting M1.0+ class flares in the next 12 hours. The first stage is a state-of-the-art
VGG-16 convolutional neural network model that outputs a flaring probability by extracting features
from raw magnetogram images. The output of this model is then used as input to an extremely randomized 
trees model in the second stage, along with various engineered physics-based and topological features 
extracted from the magnetogram. 

Our first important contribution is the impact evaluation of various dataset
manipulations and modeling strategies on the performance of a CNN-only model (i.e. the first 
stage VGG-16 model). Primary among the dataset manipulations is the dataset
augmentation. We find that performing augmentation of the minority class (using standard rotation 
and polarity swapping) on the dataset for this model yields no improvement in predictive skill.
It should be noted that, unlike some studies which incorrectly augment both training and 
testing sets, we perform augmentation only on the training set. In addition, we also study
the predictive performance of using a temporal sequence of the $B_r$ component (with proper modeling) 
as opposed to all three components --- $B_r, B_{\theta}$ and $B_{\phi}$ --- of the vector magnetogram image
to train the VGG-16 model. Our findings show that using the $B_r$ sequence is just as predictive
as the full stack, indicating the redundancy of other components. Finally, when modeling on 
a temporal sequence of image data, we show that using an LSTM layer at the end of the VGG-16 model performs
worse than using the sequence as channels to the VGG-16 input layer. 

The second focus of this paper is the use of binary categorization metrics for evaluating flare prediction models. 
While the standard TSS metric is often used, tuning hyperparameters to 
optimize on the TSS metric alone can lead to a model that highly over-forecasts, i.e. one with many 
false positives. To address this, we propose a modified alternative metric --- $\text{TSS}_{\text{scaled}}$, 
which reduces the false positive rate in optimized models.  

Our third major contribution from this paper is the use of the ERT model in the second 
stage that trains on a feature set that includes the output probability from the VGG-16 model 
from the first stage together with various engineered features. This two-stage design
offers various advantages. First, this combines the prediction power
of the automatically learned features from magnetogram images by the VGG-16 with the engineered
features shown to be skillful in flare prediction in earlier studies. This can be considered as a comprehensive way to extract as much information 
from the magnetogram data as possible. Secondly, as we show, the two-stage
model has significantly lower false positive rates compared to the VGG-16 model alone: it reduces
the false positives ($\approx$48\%) without significantly reducing the true 
positives ($\approx$12\%). Finally, the ERT model provides a way to rank the forecasting
capability of various features (VGG-16 output probability, physics-based, topological). 
In our ERT ranking, two features considerably outrank the others. The most highly ranked
is the VGG-16 output probability, indicating that the first-stage model is skillfully predictive
of flares. The second-most ranked feature is the {\tt R\_VALUE} parameter --- the total flux in the polarity inversion line --- 
a feature designed for discriminating flaring from non-flaring active regions \citep{Schrijver:2005}. 

In this work, we have explored numerous ways of extracting information from the photospheric
magnetic field for the purpose of predicting flares. As in previous studies, the overall skill of even our best model does not significantly exceed the modified climatological forecasts developed by human forecasters \citep{leka2019b}. We conclude that the information contained in photospheric magnetic field measurements alone is insufficient to predict SMEs with significantly more skill than a basic climatological prediction. In future studies we plan to  include chromospheric and coronal observations from the SDO/Atmospheric Image Assembly (AIA) instrument for training machine learning solar flare prediction models. This will be a challenging task since there are no standard AIA features (analogous to the SHARPs features for HMI) with which to form feature vector inputs. Our focus will therefore be on developing deep learning CNN models that can efficiently extract predictive information from multi-wavelength time series of AIA images. 



\section*{Acknowledgments}

This study was funded by a grant from the NASA Space Weather Science Applications Program (Grant No. 80NSSC20K1404) and by a grant from the National Science Foundation (Grant No. AGS 2001670).

\bibliography{references}


\appendix

\begin{table}[htbp]
    \centering
    \begin{tabular}{c|c|c|c|c|c|c|c|c}
        \hline
        Configuration & Optimal threshold & TPR & FPR & Accuracy & TSS & HSS & ROC AUC & PR AUC\\
        \hline 
        \hline
        $C_1$: [$B_r,B_{\phi},B_{\theta}$] & 0.4 & 0.83 & 0.02 & 0.98 & 0.81 & 0.19 & 0.967 & 0.43 \\ 
        $C_2$: $B_r$ & 0.4 & 0.84 & 0.03 & 0.97 & 0.82 & 0.16 & 0.965 & 0.43 \\ 
        $C_3$: $B_r$ stack w/LSTM & 0.4 & 0.80 & 0.04 & 0.96 & 0.76 & 0.11 & 0.975 & 0.43 \\ 
        $C_4$: $B_r$ stack as channels & 0.4 & 0.79 & 0.02 & 0.98 & 0.76 & 0.18 & 0.974 & 0.46 \\ 
        \hline
    \end{tabular}
    \caption{Performance of the VGG-16 model variants discussed in Section~\ref{sec:model}.}
    \label{tab:cnn_only}
\end{table}

\begin{table}[htbp]
    \centering
    \begin{tabular}{c|c|c|c|c|c|c|c}
        \hline
        Random seed & P & N & Architecture & TP & TN & FP & FN \\
        \hline
        \hline
        Seed 100 & 286 & 23345 & CNN-only & 217 & 21977 & 1368 & 69 \\
                           & & & CNN+ERT  & 200 & 22662 & 683 & 86 \\
        \hline
        Seed 200 & 207 & 23768 & CNN-only & 145 & 22343 & 1425 & 62 \\
                           & & & CNN+ERT & 126 & 22979 & 789 & 81 \\
        \hline
        Seed 300 & 137 & 22283 & CNN-only & 97 & 21214 & 1069 & 40 \\
                           & & & CNN+ERT & 76 & 22064 & 219 & 61 \\  
        \hline
        Seed 400 & 347 & 22760 & CNN-only & 261 & 21525 & 1235 & 86 \\
                           & & & CNN+ERT & 231 & 22015 & 745 & 116 \\  
        \hline
        Seed 500 & 228 & 22787 & CNN-only & 192 & 21117 & 1670 & 36 \\
                           & & & CNN+ERT & 172 & 21942 & 845 & 56 \\  
        \hline
        Seed 600 & 156 & 24532 & CNN-only & 108 & 22628 & 1904 & 48 \\
                           & & & CNN+ERT & 81 & 23496 & 1036 & 75 \\  
        \hline
        Seed 700 & 325 & 22187 & CNN-only & 251 & 20889 & 1298 & 74 \\
                           & & & CNN+ERT & 234 & 21395 & 792 & 91 \\  
        \hline
        Seed 800 & 217 & 24007 & CNN-only & 151 & 22360 & 1647 & 66 \\
                           & & & CNN+ERT & 139 & 22966 & 1041 & 78 \\ 
        \hline
        Seed 900 & 207 & 22425 & CNN-only & 144 & 20765 & 1660 & 63 \\
                           & & & CNN+ERT & 126 & 21697 & 728 & 81 \\
        \hline
        Seed 1000 & 184 & 23509 & CNN-only & 148 & 21994 & 1515 & 36 \\
                           & & & CNN+ERT & 146 & 22643 & 866 & 38 \\                           
                           
    \end{tabular}
    \caption{Comparison results for the CNN-Only and CNN w/ ERT architectures}
    \label{tab:comparison_raw}
\end{table}

\end{document}